\documentclass{article}
           \usepackage{graphics}
            \usepackage{natbib}
           
           \bibpunct{}{}{,}{s}{}{}
           \def\r{ $^{)}$ }
           \def\l{ $^{(}$ }
          \begin{document}
           \title{Towards a Coherent Theory of Physics and
             Mathematics: The Theory-Experiment Connection}
          \author{Paul Benioff\thanks{
           Physics Division, Argonne National Lab,
           Argonne, IL 60439; e-mail: pbenioff@anl.gov}\\[7ex]}

            \date{}
          \maketitle

           \parbox{28em}{\small
           \begin{abstract}
           The problem of how mathematics and physics are related
           at a foundational level is of much interest.  One approach
           is to work towards a coherent theory of physics and mathematics
           together. Here steps are taken in this direction by first
           examining the theory experiment connection. The role of an
           implied theory hierarchy and use of computers in comparing
           theory and experiment is described. The main idea of the paper
           is to tighten the theory experiment connection by
           bringing physical theories, as mathematical structures over $C$,
           the complex numbers, closer to what is actually done in
           experimental measurements and computations.  The method
           replaces $C$ by $C_{n}$ which is the set of pairs,
           $R_{n},I_{n}$, of $n$ figure rational numbers in some
           basis. The properties of these numbers are based on the
           type of numbers that represent measurement outcomes for
           continuous variables.

           A model of space and time based on $R_{n}$ is discussed.
           The model is scale invariant with regions of
           constant step size interrupted by exponential jumps.
           A method of taking the limit $n\rightarrow\infty$ to obtain
           locally flat continuum based space and time is outlined.
           Possibly the most interesting
           result is that $R_{n}$ based space is invariant under
           scale transformations which correspond to expansion and
           contraction of space relative to a flat background.
           Also the location of the origin, which is a
           space and time singularity, does not change under these
           transformations. Some properties of quantum mechanics, $Qm_{n}$
           based on $C_{n}$ and on $R_{n}$ space are briefly investigated.
            \end{abstract}}

            \section{Introduction}
             As is widely recognized, quantum mechanics
          and its generalizations, such as quantum field theory, is a highly
          successful theory. So far it has survived every
          experimental test. Yet in spite of this  problems
          remain. The problem of measurement is one.  Although the
          use of decoherence to solve the problem$^{(}$\cite{Zurek,Zeh}$^{)}$
          helps in that it explains the existence of the pointer
          basis in measuring apparatuses, questions still
          remain$^{(}$\cite{Adler}$^{)}$ that are related to whether quantum
          mechanics is really a theory of open systems only or
          whether there is a system such as the universe  that can
          be considered to be closed and isolated.  This is the approach
          taken by the Everett Wheeler
          interpretation$^{(}$\cite{Everett,Wheeler}$^{)}$.

          There are other open questions such as, why is
          space-time 3+1 dimensional, why are there four
          fundamental forces with the observed strengths, what is
          the reason for the observed elementary particle mass spectrum, and
          why did the big bang occur. A recent list of ten basic
          questions$^{(}$\cite{Stringth2000}$^{)}$ includes these and
          other questions. Another basic question relates to why quantum
          mechanics and its extensions is the correct
          physical theory.

          There are papers in the literature
          that address some of these questions by attempting to
          show that if things were different then life could not
          have evolved or some physical catastrophe would
          happen$^{(}$\cite{Tegmark,Vandam,Hogan,Barrow}$^{)}$. However these are all
          heuristic after-the-fact types of arguments and do not
          constitute proofs. The possibility of constructing a
          theory to explain these things, as a "Theory of
          Everything" or TOE represents a sought after goal of
          physics $^{(}$\cite{Weinberg,Greene,TegmarkTOE,Barrow}$^{)}$.

          Another very basic problem concerns the relation between
          physics and mathematics. The view that seems to be taken by most
          physicists is that the physical universe and the
          properties of physical systems exist independent of and
          a-priori to an observers use of experiments to construct
          a theory of the physical universe. In particular it is
          felt that the properties of physical systems are
          independent of the basic properties of how an observer
          acquires knowledge and constructs a physical theory of
          the universe.  This view is expressed by such phrases as
          "discovering the properties of nature" and regarding
          physics as  "a voyage of discovery".

          A similar situation exists in mathematics. Most
          mathematicians appear to implicitly accept the realist
          view. Mathematical objects have an independent,
          a priori existence independent of an observers knowledge of
          them$^{(}$\cite{Frankel,Shapiro}$^{)}$.  Much
          mathematical activity consists of discovering properties
          of these objects.

          This is perhaps the majority view, but it is not the
          only view.  Other concepts of existence include the
          formalist approach and various constructive
          approaches$^{(}$\cite{Heyting,Bishop,Beeson,Svozil}$^{)}$.
          These approaches will not
          be used here as they do not seem to take sufficient
          account of limitations imposed by physics, e. g. the
          physical nature of language$^{(}$\cite{BenLP}$^{)}$ and
          information $^{(}$\cite{Landauer}$^{)}$.

          This realist view of physics and mathematics has some problems.
          This is especially the case if  one considers
          physical systems as those that exist in and determine properties of a space-time
          framework and mathematical systems as those that exist outside of space-time and
          have nothing to do with space-time. If this is indeed the case,
          then why should mathematics be relevant or useful
          at all to physics? Also if this is the case, it is not clear how one
          acquires knowledge of mathematics$^{(}$\cite{Shapiro}$^{)}$.
          It is obvious that physics
          and mathematics are closely entwined, as is shown by extensive use of
          mathematics in theoretical physics, yet it is not clear how
          the two are related at a foundational level.

          This problem has been well known for a long time and has been much studied.
          An early and well known expression of it was by
          Wigner$^{(}$\cite{Wigner}$^{)}$ in a
          paper in 1960 entitled {\it The Unreasonable Effectiveness of Mathematics in the
          Natural Sciences}. A sampling of the extensive literature on various aspects of
          the relationship between physics and mathematics includes work on quantum
          set theory$^{(}$\cite{Finkelstein,Takeuti,Titani,Schlesinger}$^{)}$,
          the relationship between the Riemann hypothesis and aspects of quantum
          mechanics$^{(}$\cite{Odlyzko,Crandall}$^{)}$ and
          relativity$^{(}$\cite{Okubo}$^{)}$, and efforts to connect quantum
          mechanics and quantum computing with logic, languages, and
          different aspects of physics$^{(}$\cite{Ozhigov,Buhrman}$^{)}$ along
          with efforts to connect mathematical logic with
          physics$^{(}$\cite{Tegmark,Spector,Krol}$^{)}$. A related
          question that needs addressing is, Why is Physics so
          Comprehensible?$^{(}$\cite{Davies}$^{)}$.

          Other foundational issue are based on the universal
          applicability of quantum mechanics. If quantum mechanics is truly
          universally applicable, then all
          physical systems, including experimental equipment, computers,
          and even intelligent systems and language expressions are quantum systems in
          different states.  The fact that intelligent systems, and
          experimental equipment are macroscopic and follow a dynamics that,
          for most or all variables of interest, is classical mechanical does not
          exclude their following quantum dynamical laws.  Examples
          of such laws are evolution equations for density operators which are
          interacting with their environment.

          An important example of dynamical processes is a process by which an
          intelligent system compares theoretical predictions with experimental
          outcomes as part of the process of validation or refutation of a
          theory. If quantum mechanics is universally applicable,
          then both experimental apparatus and computers are
          quantum dynamical systems as is the intelligent system
          that is carrying out the validation.  In addition the
          expressions in the language of the theory from which the
          theoretical predictions are obtained are also states of
          quantum systems\l\cite{BenLP}\r.

          It follows that the process of validation (or
          refutation) of any theory
          is a quantum dynamical process described by quantum
          dynamical evolution laws. If these ideas are  applied to
          quantum mechanics itself, then one sees that quantum
          mechanics must in some sense describe its own
          validation by quantum systems. However almost nothing
          is known so far about the details of such
          a description. A possible model that may be useful for studying
          this problem is a quantum robot\l\cite{BenQR}\r as a mobile
          quantum computer interacting
          with its environment and generating output binary qubit
          states of increasing length.  One can ask if there
          exists a quantum dynamics for the robot such that the
          output binary strings  it generates can be interpreted by us, as
          external observers, as language expressions that have
          meaning to us. Can these expressions have meaning as descriptions
          of experimental tests of
          predicted properties of the robots environment? Perhaps
          more importantly one can ask if the binary states have
          meaning to the quantum robot itself, and if so, what the meaning
          is.

          The approach taken here is to work towards the
          development of a coherent theory of physics and
          mathematics that describes both physical and mathematical
          systems together in a single theory rather than as two
          separate types of systems with different theoretical
          descriptions. It is hoped that such a theory, with its
          somewhat different view of physics and mathematics, may help in
          answering some of the open questions described above.

          The next section provides a description of some aspects of a
          coherent theory of physics
          and mathematics.  The description is  brief because it is based on
          previous descriptions of this approach\l\cite{BenLP,BenTCTPM,BenRLTE}\r.
          It is presented here because it provides
          an overarching foundation or basis in which to frame and
          investigate questions of interest in this paper. However it and
          subsections \ref{IRFSS} and \ref{WC} can be skipped by readers interested
          mainly in the physical aspects discussed in this work.

          The main part of this paper is contained in Sections
          \ref{TTEC} - \ref{TLINF}.  Various aspects of the theory
          experiment connection that are of interest here are
          discussed in Section \ref{TTEC}. Included is a
          discussion of the indirectness of the status of physical
          reality as a function of the size of a system. It is
          seen that if one takes into account the requirements that equipment
          for an experiment functions properly and that this depends on the
          validity of theories on which the proper functioning is based,
          then a hierarchy of theories or parts of theories results.
          The question of why computers are
          necessary to compare theory with experiment is discussed
          from a slightly different viewpoint than that usually
          assumed. In essence the point is that computers provide
          a bridge between the type of real number names given by
          theory and the type of number generated as experimental output.

          The next two subsections discuss this connection, but
          more from the viewpoint of regarding real numbers as
          elements of equivalence classes of convergent Cauchy
          sequences of rational string numbers\footnote{One may
          also regard real numbers as infinite digit strings in
          some basis. However this requires including in a theory
          language expressions of infinite length.  This is
          excluded as language expressions are required to be of
          finite length. Rational number strings are finite digit
          strings with digits different from $0$ in both sides of
          the point separating positive and negative powers of the
          basis, as in 1101.011 in binary. Even though this does not
          include all rational numbers, rational number strings are
          dense in the rational numbers. As a result, each real number
          corresponds to a convergent Cauchy sequence of these
          string numbers.}, and measurement outcomes as
          finite digit strings in some basis,  i.e. an $n$ significant
          figure representation. The use of coarse graining to
          connect the two types of numbers is noted along with
          some problems with this connection.

          The main point of the paper is in subsection \ref{CFOMCT}.
          There it is proposed that
          the theory-experiment connection be tightened up.
          The method suggested is based on the observation that
          all physical theories to date can be described as
          different mathematical structures over $C$, the field of
          complex numbers. The method consists in replacing $C$ by
          $C_{n}$ which is a set of  finite string complex rational
          numbers of length $n$ in some basis (e.g. binary) and then
          taking the limit $n\rightarrow\infty$. In this way one
          starts with physical theories based on numbers that are much closer to
          experimental outcomes and computational numbers than are $C$ based theories.

          The rest of the paper describes some
          consequences of this replacement. Most of the discussion, in
          Section \ref{RnBST}, centers on some aspects of $R_{n}$ based space and
          time. The description of space and time, which is
          based on the properties of numbers in $R_{n}$, the real
          part of $C_{n}$, is very different from the usual
          description.  In essence it is a scale invariant
          description in that multiplication of the
          space time locations by any power of the base used in
          $R_{n}$ does not change anything. Interesting results
          include the nature of the origin as a singularity
          and the observation that scale transformations correspond to expansion
          and contraction of the space relative to a flat background
          without changing the location of the origin.

          The next two Sections, \ref{QMBCn} and  \ref{TLINF}, give
          brief descriptions of a few aspects of $Qm_{n}$, which is
          $C_{n}$ based quantum mechanics, and of taking the limit
          $n\rightarrow\infty$. The limit process is quite important
          in that one must recover continuum based space and time and
          $C$ based physics in the limit.  The paper concludes with
          a summary in Section \ref{SO}.

          The work described here represents initial steps in
          approaching a coherent theory of physics and mathematics.
          One may hope that these or related  foundational questions and the
          approach taken here are also of interest to  Asher Peres.
          His interest in foundational aspects regarding the
          interpretation of quantum mechanics\l\cite{FuPe}\r do
          overlap those of this paper.  It is a pleasure to submit this
          paper as part of a 70th birthday festschrift for Asher
          Peres.

          \section{Towards a Coherent Theory of Physics and
          Mathematics}\label{TCTPM} As noted above the goal of a coherent
          theory of physics and mathematics is to treat
          mathematical and physical systems together in a single unified
          theory rather than regarding them as completely
          distinct types of systems described by different
          theories. At present essentially nothing is known about
          the details of a coherent theory.  However one can
          summarize some basic properties that  a coherent theory should
          have\l\cite{BenTCTPM}\r.

          Probably the most important property is that it mathematics and
          physics are treated together in a single coherent theory instead
          of  treating them as separate entities as is presently done. At the
          outset one should set out what a theory is in general. The
          position taken here is based on the description provided
          by mathematical logic which is the study of theories in
          general.

          Mathematical logic makes a clear separation between
          formal theories and a universe of objects which the
          theory is supposed to describe. A formal theory $T$ is
          based on a language
          $\mathcal {L}(T)$ of expressions as strings of symbols.
          Expressions in $\mathcal {L}(T)$ are characterized in the usual
          way\l\cite{Shoenfield}\r as terms built up from function, variable, and
          constant symbols and as predicates built up from predicate symbols and
          logical connectives. A theory $T$ is created from the
          language by a choice of a set of formulas (i. e. predicates) as the axioms
          of the theory\footnote{The set of axioms is required to be
          decidable\l\cite{Shoenfield,Smullyan}\r. This is done to avoid choices
          of undecidable sets as axioms such as the set of all true statements
          for some interpretation as the axioms.}. The axioms, along with the
          logical axioms and logical rules of deduction, which are common to all
          theories, are used to define proofs as strings of formulas. A theorem
          is a terminal formula of a proof and a formula
          is a theorem if and only if it has a proof.

          It will be assumed here that a coherent theory is axiomatizable
          using first order predicate logic.  Whether this is the
          case or not is not known at present.  The usefulness of this
          assumption lies in the availability of a large body of
          knowledge for first order theories.  Also all the
          mathematics used so far by physics, and almost all
          mathematics, is covered by first order
          theories.\footnote{Second order logic quantifies over
          both individuals, as is done in first order logic, and
          over sets of individuals\l\cite{Vaananen}\r. One hopes
          that the increased strength of second order logic is not
          needed here.}

          The universe that the theory is supposed to describe
          consists of physical and mathematical systems. Hopefully
          it includes at least most or all of the physical universe
          and as much of mathematics as is needed.  Also it is not
          known how such a theory will distinguish between physical
          and mathematical systems, if such a distinction does
          exist.  However the present situation, viewing mathematical
          objects as independent of and outside space time and
          physical objects as inside of and determining the
          properties of space time, is not tenable.

          It should be emphasized that readers may disagree with
          the description of mathematical and physical universes presented
          here. Different positions include belief in the existence
          of many universes\l\cite{Deutsch,Everett,Wheeler}\r, and the belief that
          mathematical existence is different from the Platonic view. (For a recent
          review see Marek and Mycielski\l\cite{Marek}\r. From the
          viewpoint of this paper, it does not matter what the readers specific
          beliefs are about the ontology of physics and mathematics.
          It is sufficient to start from the position that mathematical and
          physical systems are different and distinct and have a different ontological
          status.

          The connection between a formal theory and its language, and a mathematical
          system or structure consisting of individuals, functions, and relations or
          properties is through an interpretation, $I,$ or map from the symbols, terms,
          and formulas of the language to the mathematical system. Variable and constant
          symbols are mapped to variables and constants in the system. Terms are  mapped
          to terms and formulas are mapped to relations in a straightforward manner that
          mirrors the inductive description of expressions in $\mathcal {L}(T)$
          that are terms and formulas. The mathematical structure for
          $T$ and $\mathcal {L}(T)$ is a  model of $T$ and $\mathcal {L}(T)$ if
          and only if all axioms of $T$ are true in the structure.  This assumes
          that $T$ is consistent (not all formulas are theorems).
          Otherwise $T$ has no models\l\cite{Shoenfield,Chang}\r.
          Note that the syntactic concept of provability  or theoremhood as a
          property of formulas in $T$ is quite separate from the semantic property
          of true or false that applies to relations in
          the mathematical structure.

          For any physical theory there is also the requirement
          that it be validated by agreement between theory and
          experiment. This extra connection, which is not part of
          mathematical theories, is essential.  It is also an
          essential connection for a coherent theory of physics
          and mathematics. If one regards a physical theory as a
          formal theory with an interpretation into some
          mathematical structure then a physical theory can be
          consistent and thus valid mathematically.  This has
          nothing to do with whether it is valid or invalid physically (i.e.
          agrees or disagrees with experiment).

          This situation is unsatisfactory in that one would like
          to bring these two notions of validity together in some
          fashion.  It is hoped that a coherent theory of physics
          and mathematics together would provide some insight into
          how this might be accomplished. It is hoped that the
          work described here in the following sections about the
          theory experiment connection might help in this
          endeavor.

          Another aspect of the relation between a formal physical
          theory $T$, the theory language  $\mathcal
          {L}(T)$, and mathematical and physical validation of $T$
          is based on the physical nature of all
          representations of language, i.e. "language is
          physical"\l\cite{BenLP}\r. Examples include written
          material, such as the text of this paper, speech,
          and trains of electromagnetic signals representing
          digitized text. Physically, printed text consists of ink
          molecules at different locations on a periodic space
          lattice and speech consists of amplitude modulations of
          a train of sound waves or of electromagnetic signals,
          such as transmission over telephone lines.
          Electromagnetic transmission of digitized text is an
          essential component of communication via the internet.

          The existence of physical representations is essential.
          Without them it would not be possible to communicate
          anything or to even think about theories. Theoretical
          explanations of what was directly experienced would not
          be possible. Note that here the emphasis here is on the
          direct physical nature of all representations of a language,
          not on a theoretical description of the physical representations
          of the language. Such a description and its relation to
          the representation is of much interest, though, as it
          leads to  questions regarding the possible existence of
          formulas that describe their own physics\l\cite{BenLP}\r.

          \section{The Theory-Experiment Connection}\label{TTEC}
          In this section some aspects of the comparison between
          theoretical predictions and experiment will be
          discussed. Much of the discussion will apply to quantum mechanics.
          Most but not all of the points discussed
          here, although known implicitly, are usually not discussed
          explicitly.  Also some novel suggestions about the
          theory experiment connection will be made.

          \subsection{Immediacy of  Reality as a Function of
          System Size} \label{IRFSS}

          An interesting aspect of the status of reality or
          objectivity for physical systems is that the directness
          or immediacy of the reality status of physical systems
          depends on the system size\l\cite{BenTCTPM,BenRLTE}\r.
          To see this one notes that the reality status, or existence, of very
          small physical systems or very large and far away systems, is more
          indirect than is the reality status of systems that are, or are
          perceived to be, moderate sized
          ($\sim$ our size) systems. For instance "This rock is
          heavy, rough and white" are immediate directly observed
          properties of a  laboratory sized system, a rock. The
          reality status of these properties is immediate and
          direct as they are direct sense impressions. No theory
          is needed to experience these properties. Another
          example is that "the sun is hot, round, and moves through the sky".
          These are directly experienced properties of a system whose
          perceived size is a few cm.. No theory is needed for
          these properties; their reality status is immediate and direct.

          The reality status of small systems such as bacteria,
          and even smaller systems such as atoms and molecules is
          more indirect in that it depends on the proper
          functioning of  equipment used to observe these systems.
          The proper functioning of the equipment depends on theory
          supported by experiment that describes how the equipment
          operates and that it performs as expected.  One concludes from
          this that the existence and other properties of
          systems, such as bacteria, atoms, and molecules, depends on
          intervening layers of theory supported by experiments.
          In this sense the reality status of these small systems
          is more indirect than that of rocks or of equipment used
          to observe the systems.

          In this sense the reality status of smaller systems such
          as protons and neutrons and the elementary particles of
          physics is even more indirect as it depends on many layers
          of intervening theory supported by experiment that
          describe the proper functioning of even more complex
          equipment.

          The same holds for large, far away cosmological systems.  The
          existence and other properties of these systems depends
          on the proper functioning of telescopes and complex
          recording equipment. This in turn depends on all the
          theory supported by experiment needed to ensure that a
          telescope and much other associated equipment functions
          properly and is not just a meaningless assembly of parts.

          The sun is a good example to show that some
          properties of a system are more indirect with more
          levels of intervening theory and experiment than for other properties.
          The properties of the sun noted above are immediate and
          direct.  The large size and distance of the sun and
          the gravitational attraction between the sun and
          planets are less direct in that they depend on
          Euclidean geometry and the Newtonian
          theory of gravity and supporting experiments.
          Observing and understanding the spectra of light
          emitted from the sun is even more indirect in
          that it is based on the theory of electromagnetism and
          the quantum mechanics of atoms and molecules and
          supporting experiments that both validate these theories
          and determine the proper function of relevant experimental
          equipment.   Finally the energy source
          of the sun as thermonuclear fusion is even more indirect
          as it depends on all the theories noted above and
          nuclear theory plus special relativity.

          These examples and many others can be described more
          abstractly by noting that the validity of an experimental
          test of a theoretical prediction depends on the proper
          functioning of each piece of equipment used in the experiment.
          But the proper functioning of each piece of
          equipment depends in turn on other supporting theory
          and experiments which in turn $\cdots$. For example, suppose an
          experiment to test the validity of a theory  uses
          two pieces of equipment, $E_{1},E_{2}$. The validity of this
          experiment as a test depends on the proper functioning of $E_{1}$
          and $E_{2}$. However, the proper functioning of $E_{1}$ also
          depends on some theory which may or may not be the same
          as the one being tested, and also on some other experiments each
          of which depend on other pieces of equipment for their
          validity. This argument then applies also  to the experiments
          used to validate the theory on which the proper functioning of
          $E_{1}$ is based. Similar statements can be made for the
          proper functioning of $E_{2}$.

          Basic examples of such equipment are those that measure time and distance.
          The truth of the assertion that a specific system, called a clock,
          measures time depends on the theory and experiments
          needed to describe the clock components and their proper
          functioning. The proper functioning of a clock
          depends on the proper functioning of each component of a
          clock. Similar arguments can be made for distance measuring equipment
          and equipment for measuring other physical parameters.

          Computations made to compare theoretical predictions with
          experiment have the same property. A computation is a sequence of
          different steps each performed by a computer which is a physical system.
          The proper functioning of the computer depends on the proper functioning
          of the many parts of the computer.  This is based on a theory, which
          may or may not be the same as the one for which the computation is
          made, and on supporting experiments that validate the theory
          on which the proper functioning of the computer and its parts is based.
          This dependence is also required to conclude that the outcome of a
          dynamical process that describes the operation of a computer is
          indeed a computation.

          This shows that the reality status of system properties
          depends on a downward descending network of theories,
          computations, and experiments.  The descent terminates at the
          level of the direct, elementary observations. These require no theory
          or experiment as they are uninterpreted and directly perceived.

          The indirectness of the reality status of systems and
          their properties is measured crudely by the depth of descent between
          the property statement of interest and the direct
          elementary, uninterpreted observations of an observer.
          This can be described very crudely as the number of layers
          of theory and experiment between the statement of
          interest and elementary observations. The dependence on
          size arises because the descent depth, or number of
          intervening layers, is larger for very
          small and very large systems than it is for moderate
          sized systems.

          The same arguments can be made for the complexity or
          indirectness of experimental
          support for a theory. The validity of each experiment as
          a test of a theory also depends on a downward descending
          network of theories, computations, and experiments that
          terminates at the level of  direct, elementary observations.
          The theory being tested by the experiment may or may not be the
          same as the theory used to support the proper
          functioning of equipment used in the experiment.

          Another measure of the indirectness of the reality of
          properties or indirectness of experimental support for a
          theory is the amount of physical resources needed to
          carry out the necessary experiments and computations. It
          is clear that the resources need to carry out
          experiments to determine properties of very small or
          very large far away systems are greater than those
          needed for moderate sized local systems.  This approach
          will not be pursued here as it is discussed
          elsewhere\l\cite{BenRLTE}\r.

          Finally it must be emphasized that the discussion is
          about the indirectness of the reality status or
          existence of systems and their properties only.  It is
          not at all suggested or implied that some systems and
          properties are more or less real than others.

          \subsection{Why Computers?}\label{WC}
          At first sight the question of why computers are needed
          for theory experiment comparison has an obvious answer:
          To compute numbers for theoretical predictions to
          compare directly with the results of experiments. Here a
          slightly different viewpoint is taken that fits in with the
          material in Section \ref{TCTPM} and with the possibility
          of a coherent theory of physics and mathematics.

          To begin with one must consider the naming or designation of real
          numbers as these are the assumed connection between
          theory and experiment. That is, in quantum mechanics at least,
          all predictions of theory are real numbers. This follows from
          the basic mathematical structure of quantum mechanics.
          This is the case even for systems in discrete states such as
          spin projections. Here the eigenvalues of the Pauli operators
          such as $\pm 1$ for $\sigma_{z}$ are integers only through
          the inverse of an obvious map of integers to real numbers,
          $\pm 1 \mapsto \pm 1.0000\cdots$.

          The physical nature of language requires that real number names
          must be strings of symbols. One possibility that comes to mind
          is to extend any $k-ary$ representation used for the rational
          numbers to include infinitely long symbol strings. If this
          were possible, then each real number would have a name. The problem
          is that such strings are excluded on physical grounds in that
          all symbol strings  must be finite in length.

          An immediate consequence of this requirement is that
          almost all real numbers cannot be named as there are
          only  a countable infinity of finite symbol strings.
          Thus one needs to examine how names of real numbers can
          be named under this limitation.

          In general  there are several ways to name some real numbers.
          For a $k-ary$ representation one type of name is
          $\underline{a}*(\underline{b})^{\infty}$.
          Here $\underline{a}$ and $\underline{b}$ denote finite
          sequences of $k$ digits and the superscript denotes
          infinite repetition of $\underline{b}$.  The symbol $*$ denotes concatenation.
          The location of the $"k=al"$ point, which is arbitrary, denotes
          the magnitude of the number.

          As is well known, this type of naming gives names to the real number
          equivalents of all rational numbers.  Another type of naming corresponds to
          replacing an infinite $k-ary$ sequence by a computation procedure
          that can compute
          successive approximations to the infinite sequences. Such a
          procedure might, for example, compute for each $n$ the
          first $n$ binary digits of an infinite string. Since
          there are at most countably many computation procedures
          under any given representation of computability, e.g. as
          Turing machines\l\cite{Rogers}\r,  there are at most
          countably many computable real numbers.  Here a real
          number is computable if there is a Turing machine that
          can compute successive approximations to the number.  In
          this case, computable real numbers can be given a
          (natural number) name corresponding to the position, in
          a listing of all Turing machines, of any Turing machine
          that approximates the number.

          This method of naming gives many names to each
          computable real number as there are many possible ways
          to compute successive approximations to a given number.
          It includes the real number representations of
          rational numbers as well as real numbers such as $e,\pi,
          \sqrt{2},$  and many others expressible as the limit of a
          computable convergent series.

          Another method of naming real numbers is based
          ultimately on the property that each polynomial equation
          of odd degree has a real number solution. Algebraic real
          numbers are a special case of this in that the
          coefficients of the polynomials are integers. Since all
          integers (and rational numbers) have names, each algebraic
          real number $r$ can be named by
          a polynomial equation, $\sum_{j=0}^{n}a_{j}x^{j}=0$
          where each $a_{j}$ is an integer,  which has $r$ as a
          solution. If the equation has many solutions, then naming
          requires an additional specification, such as that based
          on use of an ordering (e.g. the $nth$ smallest solution).

          Physical theories greatly extend this method of naming
          real numbers in that all theoretical predictions are
          expressed as equations which supposedly have real number
          solutions. This is the case for any  physical
          theory that describes a mathematical structure over a
          field of complex numbers.  Quantum mechanics and its
          generalizations, such as quantum field
          theory, QED, and QCD are prime examples with this structure. So are
          special and general relativity. In these cases predictions
          correspond to theorems as statements in the theory language
          that define unique real numbers that are solutions of equations
          described in the theorem. Note that each coefficient appearing in
          these equations must be a name of some specific element representing
          a dimensional or dimensionless physical quantity in the mathematical
          structure of the theory.

          To see this in more detail it is worth looking at a specific
          example in quantum mechanics.  Suppose one is interested
          in the low lying atomic energy levels of an atom of mass $M$ with
          $Z$ electrons.  The prediction of the ground state energy of the
          of the atom as the lowest eigenvalue of the Hamiltonian $\tilde{H}$
          is obtained by solving the time independent Schr\"{o}dinger
          equation $\tilde{H}\Psi-E\Psi =0$ for the energy of the ground state.
          The Hamiltonian has the well known form in the space
          coordinate representation\l\cite{Pines}\r,
          \begin{equation}\label{Hatom}
          \tilde{H}=\sum_{j=1}^{Z}(-\frac{\hbar^{2}}{2m}
          \underline{\nabla}^{2}_{j})
          -\frac{\hbar^{2}}{2M}\underline{\nabla}^{2}_{0}
          +\sum_{1\leq i<j\leq Z}\frac{e^{2}}{|r_{i}-r_{j}|}
          -\sum_{j=1}^{Z}\frac{Ze^{2}}{|r_{j}-r_{0}|}
          \end{equation}where $r_{0}$ is the nuclear coordinate.

          $\tilde{H}$ is specified in the sense that $\hbar,e,m,M$ are
          names for specific physical parameters with specific
          real number values.  The presence of space
          variables $r_{j}$ and the state $\Psi$ is taken care
          of by integration over the space variables and requiring
          that $\Psi$ is the ground state. In more formal
          detail the prediction is in terms of a theoretical
          statement about certain real numbers that must be a theorem
          of quantum mechanics: $$\exists E\exists\Psi\{
          E=(\Psi,\tilde{H}\Psi)\bigwedge [(\Psi,(\tilde{H}-E)^{2}
          \Psi)=0]\bigwedge\forall E^{\prime}( E^{\prime}
          =(\Psi,\tilde{H}\Psi)\Rightarrow E\leq  E^{\prime})\}.$$
          This statement says that there exists a real number $E$ and a
          state $\Psi$ where $E=(\Psi,\tilde{H}\Psi)$ and $\Psi$ is
          dispersion free for $\tilde{H}$ (an eigenstate) and $E$
          is the smallest  value of $(\Psi,\tilde{H}\Psi).$

          At this point one realizes there is a disconnect between
          theory and experiment.  Theory provides a name of a real
          number as a solution of an equation with auxiliary
          conditions to specify which solution, if necessary. However
          experiment gives a finite string of digits in some representation
          (binary, etc.) that represents a number. This follows from
          the observation that each experiment yields an apparatus in a
          final state, part of which is interpreted as a finite digit string.

          This is the case for both digital and analog outcomes since for
          analog outcomes one must read a pointer position on a
          dial relative to background marks. The fact that the
          string is necessarily finite is accounted for by saying
          that the measurement outcome is an "approximation" of
          the true value, or that it is a consequence of the
          finite precision and accuracy of the measurement
          procedure.

          There are two problems here: One is that the measurement
          outcome is an approximation of a real number and is not
          a real number. The other follows from the observation
          that the measurement outcome, as a finite digit string,
          is considered to be an initial string of an infinite digit
          string that corresponds to the outcome in the ideal limit
          of infinite precision and accuracy. However this is
          quite a different type of name than that provided by
          theory which names a real number by an equation for
          which it is a solution.

          This disconnect is bridged by the use of computers.
          Here the purpose of a computer is to convert or
          translate a real number name as a solution of an
          equation to a finite digit string that is an
          approximation to the real number solution.  This digit
          string can be directly compared to the measurement
          outcome to look for agreement or disagreement.

          As is well known much effort in physics goes into
          computing approximate solutions to theory equations that
          are difficult to solve. Here consideration will be
          limited briefly to two aspects.  One is that, as a
          physical dynamical operation with physical systems, a computation
          is based on arithmetic operations on
          finite digit strings, This arithmetic, called computer
          arithmetic, is much studied\l\cite{Comparith}\r.
          Differences between computer arithmetic and real number
          arithmetic include the use of roundoff computations
          to minimize the effects of working with finite digit strings.
          The computation output is also a finite digit string that
          is supposed to be an approximation to an infinite digit
          string real number.

          The other aspect relates to the correspondence between
          the equation to be solved and a particular computer. If
          the standard Turing machine representation is used the
          correspondence associates one or more Turing machine
          names with the equation to be solved. The same holds for
          any other standard representation.

          The main point to make here is that the determination
          of this map or correspondence is often nontrivial.  In
          any case, the details of the map do not seem to be part
          of the physical theory.  Instead the map seems to be
          part of the metatheory and language used to describe the
          physical theory and its mathematics.  A goal, which is
          open at present, is to extend the theory so that it can
          at least describe and may possibly prove which
          computations (Turing machine names) will solve a
          specified equation to a specified approximation.

          \subsection{Real Numbers and Outcomes of Measurements
          and Computations}\label{RNOMC}
          The goal here is to examine the relation
          between outcomes of measurements and computations, and
          real numbers. It will be seen that there are problems
          that do not seem to be adequately treated so far.

          To begin it is to be emphasized that the numbers of interest
          in a theory such a quantum mechanics are real
          and complex numbers.  This is based on the mathematical
          structure of quantum mechanics as state spaces, operator
          algebras. and other mathematical systems all based on the field
          $C$ of complex numbers. The elements of $C$ are often shown as
          pairs $(r,i)$ of real numbers denoting the real and imaginary
          components. Real numbers can also be described as a
          subfield $R$ of $C.$ Rational numbers are
          present as a  subfield $Ra$ of $R,$ integers are a
          subring of $Ra$, and the natural numbers are the
          nonnegative integers.

          As is well known, rational numbers, integers, and natural
          numbers are present as special types of real
          numbers.  In a $k-ary$ representation  each rational number
          is represented as an infinite string of $k-ary$ digits
          whose tail is an infinite repetition of some finite digit
          string. Rational string numbers are a restriction of
          rational real numbers to strings whose tail is an infinite
          string of zeros. Integers, and natural numbers have a similar
          representation where the tail of zeros begins at the "$k-al$" point.
          Binary examples of the four types are respectively,
          $-11001.01(011)(011)(011)\cdots,\;1011.110110000\cdots,$
          $-11101.000\cdots,\;1101.000\cdots.$

          However outcomes of measurements and of computations are,
          physically, states of parts of a system. For computers or
          measurement equipment they are states of some output display
          such as a liquid crystal display or an output register.
          Quantum mechanically the states of these output parts can
          be represented by  density operators obtained by
          tracing over other system degrees of freedom
          including the environment if needed.

          As is well known these output states of computations or
          measurements are interpreted as  numbers represented by
          finite strings of digits or, using the language of information
          theory, as product states of finite strings of kits or qukits
          ($k-ary$ representation). The fact that these strings
          are finite is based in part on the physical limitation
          that infinite output strings are impossible to create or
          read in a finite amount of space and time. Also equally well
          known are the finite accuracy and precision  associated with all
          computations and measurements. This is often expressed by referring
          to measurement outcomes as good to $n$ significant figures.

          The problem of how one relates computation or
          measurement outcomes as finite strings to real numbers
          is well known.  This is especially
          so for measurement of observables with continuous
          spectra such as position and momentum.

          One approach to the problem is the use of coarse graining.
          In this case assume that the computer or measurement equipment
          has $n$ binary significant figure outputs over some range
          from $a$ to $b$. Coarse graining assumes that each such
          output represents  a bin or range of real numbers.  For
          example for $n=5$ the output $11.011$ might represent
          all real numbers $r$ where $11.0101000\cdots\leq r \leq
          11.0111000\cdots$.

          Of course many different binnings are possible. More
          generally if the output is represented by
          $\underline{s}.\underline{t} $(fixed point
          representation) the number of significant figures is
          given by $L(\underline{s})+L(\underline{t})$ where
          $\underline{s}$ and $\underline{t}$ are binary strings
          of respective lengths  $L(\underline{s})$ and
          $L(\underline{t}).$ Then one can represent a coarse
          graining by the choice of two real numbers $\Delta_{l}$
          and $\Delta_{u}$ where the output $\underline{s}.\underline{t}$
          represents all real numbers $r$ where
          $\underline{s}.\underline{t}000\cdots -
          \Delta_{l}\leq r\leq \underline{s}.\underline{t}000\cdots +
          \Delta_{u}.$ Many choices are possible for the values of
          $\Delta_{l}$
          and $\Delta_{l}.$  They can be set independent of
          $\underline{s}.\underline{t}$, or they may depend on
          $\underline{s}.\underline{t}.$  Other types of maps are
          statistical with a chosen probability distribution over
          different binnings.

         This exposes the problem of how one relates an $n$ significant
         figure output of a computation or measurement
          to the real numbers.  Some type of map is clearly needed for
          comparison of theory with experiment for any theory as a
          mathematical structure over $C$, such as quantum mechanics.
          The map should apply for arbitrary values of $n$ as
          different experimental and computational setups can give
          different values of $n$.

          This is not a new problem.  Recent work to treat this
          problem\l\cite{Meyer1}\r separates the Hilbert space of quantum
          mechanics into a product space ${\mathcal H}_{coarse}
          \otimes{\mathcal H}_{fine}$
          where states in ${\mathcal H}_{fine}$ are not observable. Other
          work\l\cite{Meyer2}\r is based on the use of state vectors with
          rational components. This has been criticized as being inconsistent
          with axioms of geometry\l\cite{Peres}\r.

          It should be emphasized that for essentially all experiments done
          to date, which coarse graining is chosen does not affect the
          comparison of theory and experiment. If it did this
          would have been discovered by now.  However the method of
          connecting $n$ significant figure outputs of measurements
          and computations to theory is a question of some concern,
          particularly if one is interested in foundational aspects
          of the relationship between physics and mathematics.

          \subsection{Connecting $n$ Figure Outcomes of
          Measurement and Computation to Theory}\label{CFOMCT}

          An important aspect of a coherent theory of physics and mathematics
          is that it be very closely connected to computations and experiment.
          One method to consider is to tie the theory in to what is actually
          done in experiments and computation.  The method proposed
          here is to replace a physical theory $T$, as a mathematical
          structure based on $C$ by a theory $Th_{n}$ based on $C_{n}$.

          The idea is to let $C_{n}$ be a set of numbers with associated
          arithmetic relations that more closely ties in with the properties of
          computation and measurement outcomes than $C$ does.  In this case
          experimental and computational support for $Th_{n}$ would be
          based on computations and measurements with outcomes in  $C_{n}$.
          In essence these are numbers given to at most $n$
          significant figures in a $k-ary$
          representation for a fixed $k$ value. As such they more closely represent
          measurement and computational outcomes than do the numbers in $C$.

          Measurement and computations with $m$ significant figure outcomes are
          included in $Th_{n}$ if $m\leq n$. In this case a coarse graining would be
          needed to map the $m$ significant figure outcomes into $C_{n}.$
          However such a coarse graining, which is from $m$ significant figure
          numbers to $n$ significant figure numbers, is easier to accept than
          one into $C.$ Also the numbers in $C_{n}$ and the outcome numbers are
          of the same type in that they are both finite length string numbers..

          Measurement and computation outcomes with $m>n$ are not in the domain of
          $Th_{n}$ as the outcomes are not in $C_{n}.$ These belong to all theories
          $Th_{\ell}$ where $\ell\geq m.$  This shows that one must consider the
          sequence of theories $Th_{n}$ with increasing $n$ along with
          the limit $\lim_{n\rightarrow \infty}Th_{n}$. How the limit
          is taken is a matter of great importance.  It would be expected that
          the limiting process is such that the limit theory includes
          much, or almost all, of the existing $C$ based physical theory. This is
          necessary because of the great success that $C$ based physical theories
          have had to date. Also this approach may shed new light on foundational
          aspects and help solve basic open
          problems in physics\l\cite{Stringth2000}\r.

          There are several candidate definitions for the numbers
          in $C_{n}.$  One is based
          on the floating point arithmetic used in computations. Each complex number
          in $C_{n}$ is a pair of string number pairs,
          $\{(\pm \underline{s}_{r},\pm \underline{e}_{r}),(\pm
          \underline{s}_{i},\pm\underline{e}_{i})\}.$ The number $0$ is included
          as $(\underline{0}_{[1,n]},\underline{0}).$ The subscripts $r,i$
          denote the real and imaginary components. For a $k-ary$ representation
          $\underline{s}$ denotes a string of $n$ $k$ basis digits where the leftmost
          digit $\underline{s}_{1}\neq 0$ and the "$k-al$" point follows the
          righthand most digit, $\underline{s}_{n}$. The subscripted brackets $[a,b]$
          denote place labels ranging from $a$ to $b$. Here $\pm\underline{e}$
          denotes a $k$ basis representation of integral powers of $k$.

          As an $n$ significant figure representation of numbers,
          these string pairs correspond
          to the number representations used by computers both during computation and
          as input and output.  For example all computers are based on $n$ bit
          precision arithmetic and treat numbers within some very large range.
          This is shown by the IEEE floating point arithmetic standards used by most
          computers. For example some computers use $32$ bit single precision
          arithmetic on all numbers with magnitude $m$
          between $10^{+30}$ and $10^{-30}$,
          or $64$ bit double precision arithmetic on all numbers between
          $10^{+300}$ and $10^{-300}$.

          A  slightly different, but equivalent form of number is used
          to report experimental values of physical quantities. These are
          often given as $\underline{s}_{1}.\underline{s}_{[2,n]}\times
          10^{\underline{e}}$ where $\underline{s}_{1}\neq 0.$  Often an uncertainty
          $\pm 0.\underline{0}_{[2,n-1]}a_{n}a_{n+1}\times 10^{\underline{e}}$ is
          associated with the value.

          The importance of $n$ significant figure numbers to experiments is also
          shown by the observation that in general the magnitude of the
          error of a measurement is proportional to the magnitude of
          the quantity being measured. Even more important
          is the fact that the proportionality factor is almost independent
          of the type and magnitude of the measured parameter. In the language
          used here this means that for most experimental values of physical quantities,
          the value of $n$ is almost independent of the magnitude of the quantity being
          measured.  For example, for most distance measurements $2\leq n\leq \sim 10$ for
          distance measurements over a range of $10^{31}$, from nanometers to
          millions of light years.

          The type of representation chosen here is based on outcomes of continuous
          value parameter measurements such as those for space or momentum. These
          are recorded on a set of $n$ dials or registers  where each dial or
          register records the digits $0-9$ (decimal).
          The readings $\underline{0}_{[1,n]},$
          $\underline{0}_{[1,n-1]}1_{n},$ and
          $\underline{9}_{[1,n]}$ correspond to nondetect,
          threshold detection (at
          the sensitivity level of the equipment), and the
          maximum value possible for the
          equipment used. This type of equipment is in wide
          use in flow measuring meters as
          in utility electric and gas meters.

          This describes measurement outcomes for just one
          apparatus over a finite range from nondetect to detect
          to maximum.  The whole range of quantity values from $0$
          to $\infty$ can be covered by an infinite hierarchy of
          measurement apparatuses where just off scale for the
          $jth$ apparatus corresponds to detection for the
          $j+1st.$  The hierarchy, which is infinite in both directions,
          can be represented schematically in
          binary as $$\begin{array}{ccccc}\infty\cdots & j+1 & j
          & j-1& \cdots -\infty
          \\\mbox{} & \left\{\begin{array}{c}\mbox{nondetect} \\
          \underline{0}_{[1,n]}.
          \end{array}\right \} & \underline{1}_{[1,n]}. & \mbox{off
          scale}&\mbox{} \\\mbox{}&\underline{0}_{[1,n-1]}1_{n}. &\left
          \{\begin{array}{c}\mbox{off scale} \\\underline{1}_{[1,n]}.+
          \underline{0}_{[1,n-1]}1_{n}.\end{array}\right \} & \mbox{offscale}&
          \mbox{} \\\mbox{}&\mbox{nondetect} & \mbox{nondetect} &
          \underline{1}_{[1,n]}.&\mbox{}\end{array}$$

          The hierarchy suggests a representation in the form
          $(\pm \underline{s}_{[1,n]}.,
          \pm\underline{ne})$ where $\underline{s}_{[1,n]}.\neq
          \underline{0}_{[1,n]}.$ if
          $\underline{e}\neq \underline{0}$ and $\underline{e}$ is any integer.
          Note that the exponent $\underline{ne}=n\underline{e}$
          depends on $n$ and  $\underline{e}$.  Here the use of significant figure is
          different from usual in that all $0s$ to the left of nonzeros,
          as in $0010$, are
          significant. This is based on the significance of the
          leftmost $0s$ in this type of measurement. The "bi-nal"
          point to the right is also to be noted.

          For reasons that will become clear later on, a symmetric
          form of these numbers is much better.  This form
          corresponds to the strings
          \begin{equation}\label{symnos}
          (\pm\underline{s^{\prime}}_{[1,m]}.,m(\underline{e}-\frac{1}{2}))=
          (\pm\underline{s}_{[1,n]}.\underline{t}_{[1,n]},
          2n\underline{e})\end{equation}
          Here $m=2n$ and $\underline{s^{\prime}}_{[1,m]}=
          \underline{s}_{[1,n]}*\underline{t}_{[1,n]}$ where $*$
          denotes concatenation. This representation is valid for
          any basis $k$.  Here emphasis is placed on a binary basis
          $k=2$ although sometimes a decimal basis is used for
          illustration.

          The ordering of these numbers can be expressed as an infinite
          alternating sequence of $2^{2n}-1$ numbers with constant spacing equal
          to $1\times2^{-2n(\underline{e}-1/2)}$. These regions are
          separated by exponential jumps of $2^{2n}$, where $\underline{e}
          \rightarrow\underline{e\pm 1}$. Note that there is no least or greatest
          number. Here is an example for $n=2$.$$\begin{array}{l}
           \hspace{1.6in}\vdots
          \\ 00.01,00.10,\cdots,10.00,10.01,\cdots,11.11 \;\times
          2^{4(\underline{e}+1)}\\
          00.01,00.10,\cdots,10.00,10.01,\cdots,11.11\;\times
          2^{4\underline{e}}\\00.01,00.10,\cdots,10.00,10.01,\cdots,11.11
          \;\times 2^{4(\underline{e}-1)} \\ \hspace{1.6in}\vdots\end{array}$$

          Arithmetic with these numbers is is interesting as it combines both
          region and jump arithmetic. Region arithmetic,
          which applies in the regions with
          $2^{2n}-1$ steps of constant spacing, is the
          same as computer arithmetic except
          that the range over which the arithmetic applies is
          proportional to string length instead of being
          independent of it. An example would be $n$ bit arithmetic over a range of
          $10^{\pm nc}$ where $c$ is some constant.

          Jump arithmetic consists of arithmetic operations between
          numbers in different regions or which take numbers in one region to
          those in another. It also depends on the significance of leading
          or trailing $0$ strings. Here are some examples for $m=2n=4$.
          $$\begin{array}{l}00.10\times 2^{8}+_{4}00.10\times
          2^{4}=_{4}00.10\times 2^{8}
          \\10.10\times 2^{4}+_{4}11.10\times 2^{4}=_{4}00.01\times
          2^{8}\\11.10\times 2^{8}\times_{4}10.01\times 2^{4}=_{4}00.01\times
          2^{16}\end{array}$$ Extensive roundoff has been used in these examples.
          $2n=4$ has been used for the subscripts $m$ on $=_{m},\;
          +_{m},\;\times_{m}.$
          String $ab.cd=_{4}a^{\prime}b^{\prime}.c^{\prime}d^{\prime}$ if the
          corresponding primed and unprimed digits are the same.

          The reader may wonder why one has to contend with both region and jump
          arithmetic when computer arithmetic used in physics is limited to the
          region type. The reason is that for low values of
          $n$ extensive roundoff is needed and arithmetic operations often take
          numbers from one region to another.\footnote{Consider for example
          binary arithmetic as done on a computer based on
          $4$ bit strings.} As $n$ increases
          region arithmetic increasingly dominates.  For large $n$
          the need for jump arithmetic decreases until it
          disappears in the limit $n\rightarrow\infty$. The reason
          for this will be seen later on.

          Here the set of complex string numbers $C_{n} =
          \{R_{n},I_{n}\}$ where $R_{n}$ and
          $I_{n}$ each consist of the set of all  string numbers in the form
          shown by Eq. \ref{symnos}. There is no bound on the magnitude of the
          numbers in $R_{n}$ (and $I_{n}$) as the length, $L(\underline{e})$, of
          the exponent is unbounded.

          Before applying these ideas to any theory based on these numbers
          it is worth emphasizing several aspects. One is that
          mathematical aspects of the theory are based on the
          numbers in $C_{n}$.  No other numbers are available
          for the theory.  For instance, arithmetic equality for two
          numbers $(\pm\underline{s}_{[1,n]}.\underline{t}_{[1,n]},
          \pm 2n\underline{e})$ and
          $(\pm\underline{s^{\prime}}_{[1,n]}.\underline{t^{\prime}}_{[1,n]},
          \pm 2n\underline{e^{\prime}})$
         expressed by $$(\pm\underline{s}_{[1,n]}.\underline{t}_{[1,n]},
         \pm 2n\underline{e}) =_{2n}
         (\pm\underline{s^{\prime}}_{[1,n]}.\underline{t^{\prime}}_{[1,n]},\pm
         2n\underline{e^{\prime}}),$$ holds if
         the signs are the same and $\underline{s}=
         \underline{s^{\prime}}$, $\underline{t}=
         \underline{t^{\prime}}$, and $\underline{e}=
         \underline{e^{\prime}}.$  Note the different uses of equality.
         The arithmetic definition, $=_{2n},$ depends on $n$. Two
         strings are equal if they are identical as
         strings.

         All the other arithmetic operations are also subscripted
         by $2n$ as they depend on $n.$ Examples are (subscripts
         $[1,n]$ are suppressed)
         $$(-\underline{s}.\underline{t},2n\underline{e}) +_{2n}
         (+\underline{s^{\prime}}.\underline{t^{\prime}},
         2n\underline{e^{\prime}})=_{2n}
         (+\underline{s^{\prime\prime}}\underline{t^{\prime\prime}},
         2n\underline{e^{\prime\prime}})$$ and $$(-\underline{s}.
         \underline{t},2n\underline{e})
         \times_{2n}(+\underline{s^{\prime}}.\underline{t^{\prime}},
         2n\underline{e^{\prime}})
         =_{2n}(+\underline{s^{\prime\prime}}\underline{t^{\prime\prime}},
         2n\underline{e^{\prime\prime}}).$$ Here  $+_{2n}$
         and $\times_{2n}$ include
         roundoff operations in their definitions.

         The other aspect of $Th_{n}$ is that, at first sight,
         one can think of many reasons why theories based on
         $C_{n}$ instead of the usual complex number continuum
         $C=C_{\infty}$ are not useful.  Examples include the loss
         of closure or completeness properties for components
         of the  mathematical structure over $C_{n}$, the
         explicit $k$ dependence of the definitions of $=_{n}$ and the
         arithmetic operations,  and the specific method
         of roundoff. Also because of roundoff, addition is not
         associative.

         As is well known from the operations of computers based
         on computer arithmetic, these aspects do not cause
         problems for almost all operations. For axiomatizable
         properties such as associativity, the usual axiom of
         unconditional associativity would be replaced with a
         conditional statement that if numbers in a triple have
         certain specific properties, then addition ($+_{n}$) is associative.

         What is important, though, is that  these undesirable
         properties disappear in some fashion as $n\rightarrow
         \infty.$ How this happens depends on the property and how
         the limit is taken. For example, the importance of roundoff
         decreases as the length of bit strings used in a computation
         increases relative to the length of the input and output
         strings. Possible meanings of disappearance
         include the idea that the probability of states appearing
         in the theory experiment connection with one or more of
         these undesirable properties goes to $0$ as $n\rightarrow
         \infty.$ Details on this must await additional work especially
         on how the approach of $n$ to $\infty$ is taken.

        \section{$R_{n}$ Based Space and Time}
         \label{RnBST}

         The structure of $R_{n}$ space and time  is  different
         from that  based on $R.$  The reason is that
         the only numbers available are those in $R_{n}$.
         The numbers in $R_{n}$ are taken here to have the form
         $(\pm \underline{s}_{[1,n]}.\underline{t}_{[1,n]},
         \pm\underline{2ne})$
         where $\underline{s}_{[1,n]}.\underline{t}_{[1,n]}\neq
         \underline{0}_{[1,n]}.\underline{0}_{[1,n]}.$
         This form (Subsection \ref{CFOMCT}) will be used as
         it has the right limit properties. The number $0$ is added as the
         string $(\underline{0}_{[1,n]}.\underline{0}_{[1,n]},0).$

         Two aspects need emphasis. The properties of space and
         time, as shown in the figures, look very
         different from the usual continuum space and time. This is
         a consequence of the facts that the differences are greatest
         for small values of $n$, and it is possible to
         create illustrative figures for small values of $n$ only.
         Also one must keep in mind that the figures are drawn on
         a usual continuum based background.  The background space
         is not part of space or time in a theory such as $Qm_{n}$. It is instead
         included in $C$ based physics, such as $Qm=Qm_{\infty}$.
         This emphasizes the importance of taking the limit
         $n\rightarrow\infty$. The fact that $C_{n}$ based space
         and time looks strange for finite $n$ is not important.  What is
         important is that a continuum space and time that
         is locally flat be recovered
         in the limit $n\rightarrow\infty.$

         In what follows, the $\pm$ sign on $\underline{2ne}$ is
         assumed to be included in $\underline{e}.$ Thus
         $\underline{e}$ is a $k-ary$ string representing an
         integer and $\underline{2ne}=2n\times\underline{e}$ represents
         ordinary integer multiplication. The forms
         $\underline{2ne}$ or $2n\underline{e}$ are
         interchangeable as they are equivalent.

\begin{figure}[t]
            \begin{center}
          \resizebox{100pt}{100pt}{\includegraphics[300pt,270pt]
          [520pt,490pt]{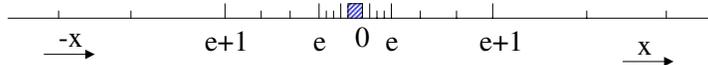}}\end{center}
          \caption{Coordinates for one Space Dimension for
          Positive and Negative $x$
          Values for $k=2n=2$.  The jump locations are shown by
          $e$ and $e+1.$ The crowding
          towards  $x=0$ is shown by  parallel line shading.}\label{1}
          \end{figure}

         The properties of $R_{n}$ based space can be seen by showing the
         point locations in one space dimension.  This is shown in
         Figure \ref{1} for $n=1$ and $k=2$ (binary)
         for positive and negative
         values of $x.$  The allowed point locations are shown
         by the vertical ticks along the line.

         Probably the most interesting aspect of the figure is the crowding
         together of the points as one approaches the $x=0$ point. The
         rate of crowding or decrease in neighborhood spacing
         can be best described as equal spacing punctuated by exponential
         decreases or increases as one moves toward
         or away from the origin at $0$ which is an accumulation
         point. The shading at $0$ represents
         the crowding of the points.  As is shown in
         the figure for $2n=2=k$, for each
         $e$ there are $2^{2n}-1$ intervals of
         spacing $2^{2n(e-1/2)}$. Then the interval
         spacing changes to $2^{2n(e-1/2\pm 1)}$ for the next $2^{2n}-1$
         intervals. As $e$ is not bounded
         there is no lower or upper bound on the interval size.

          This constancy of neighboring point distances punctuated by
         exponential jumps holds in $R_{n}$ for any value of $n$ and $k$. For
         each value of $\underline{e}$, there are $k^{2n}-1$
         successive locations each separated by a distance of
         $k^{2n(\underline{e}-1/2)}.$
         The separation distance jumps to $k^{2n(\underline{e}+1/2)}$ as
         $\underline{s}_{[1,n]}.\underline{t}_{[1,n]}$ goes from
         $\underline{1}_{[1,n]}.\underline{t}_{[1,n]}$ to
         $\underline{0}_{[1,n]}.\underline{0}_{[1,n-1]}1_{n}$
         In the opposite direction the separation distance drops to
         $k^{2n(\underline{e}-3/2)}$ as one goes from $\underline{0}_{[1,n]}.
         \underline{0}_{[1,n-1]}1_{n}$
         to  $\underline{1}_{[1,n]}.\underline{1}_{[1,n]}$.

         The locations in Fig. \ref{1} have the property that they
         are scale invariant under multiplication by
         an integral power of $2^{2n}$. The
         multiplication is in effect an exponent
         translation by replacing $e$ by $e+j$
         where $j$ is any integer. This is quite
         different than the usual lattice used
         in physics (e.g. lattice gauge theory).
         There the system is invariant
         under addition of an integral number of unit intervals.

\begin{figure}[h]
            \begin{center}\vspace*{3.5cm}
          \resizebox{100pt}{100pt}{\includegraphics[300pt,100pt]
          [520pt,320pt]{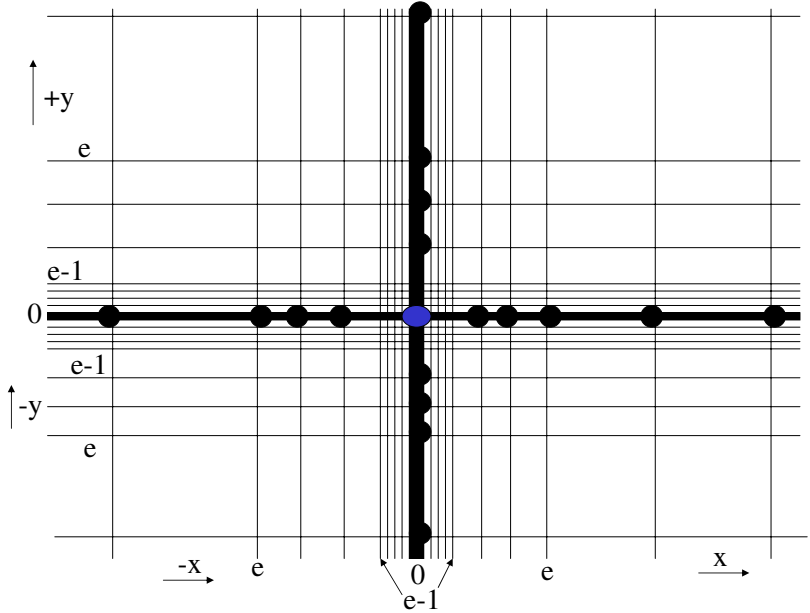}}\end{center}
          \caption{Coordinates for two Space Dimensions for all
          $R_{n}$ Values of $x$ and  $y$.
          The intersections of the grid lines show the allowed
          space locations. The jump
          locations are given by $e$ and $e-1.$ The crowding
          towards the $x=0$ and $y=0$ axes is shown by
          parallel line shading. Circles on the axes
          and at the origin denote the points of accumulation.}\label{2}
          \end{figure}

         Figure \ref{2} extends Fig \ref{1} to two space dimensions in
         Cartesian coordinates. The space point locations, which are discrete,
         correspond to the intersections points of lines parallel to the
         $x$ and $y$ axes. The crowding of the lines towards the axes
         shows that points on the axes are accumulation points.  These
         are denoted by circles on the axes.

            The figure shows that for each value of $y$, $x=0$ is an accumulation
         point from either $x$ direction. Similarly for each value of $x,$
         $y=0$ is an accumulation point for $y$ values
         from either direction. The origin is an accumulation
         point for both $x$ and $y$ values.

         Extension of this description to three dimensions is
         straightforward. There one has three types of accumulation
         points,  the origin which is a three dimensional point of
         accumulation, points on the $x,y,$ and $z$ axes which are
         two dimensional accumulation points, and points on the
         $x-y,x-z,$ and $y-x$ planes which are one dimensional.

         The proof of this singular nature of the points of
         accumulation rests on the observation that, unlike other points, they do not
         have nearest neighbors in one or more dimensions. The
         origin has no nearest neighbors in any dimension. The
         points on the axes have nearest neighbors in one
         direction and not in the other.  Points on the
         coordinate planes in three dimensional space have nearest
         neighbors in two dimensions but not in the third.

         As is well known ordinary flat space and the usual
         lattice representations are invariant under translations.
         Scale transformations play the same role for the space
         described here in that it is invariant under scale
         "translations". For the representation used here for each
         value of $k$ and $n$ the scale transformation corresponds
         to any translation of the exponent $\underline{e}$ in the
         number $(\underline{s}_{[1,n]}.\underline{t}_{[1,n]},
         2n\underline{e})$. This
         corresponds to replacing $\underline{e}$ by
         $\underline{e}+j$ where $j$ is any integer and gives a
         scale change by a factor of $2^{2nj}.$

\begin{figure}
            \begin{center}
          \resizebox{100pt}{100pt}{\includegraphics[300pt,140pt]
         [520pt,370pt]{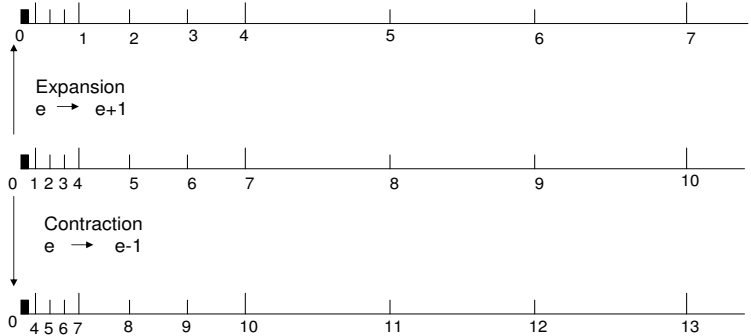}}
          \end{center}
          \caption{An example of scale
          transformations showing expansion $j=+1$ (top figure) and
          compression $j=-1$ (bottom figure) in one dimension for
          $k=2n=2$ and positive values. The
          numbers under the tick marks show the effect of the transformations
          on individual points. The origin at the short shaded
          rectangle is left unchanged
          with the region around the origin serving as an
          infinite source of points
          for expansion and an infinite sink for compression. The
          corresponding figure for
          negative values is obtained by reflection
          through the origin.}\label{3}
          \end{figure}

         The effect of this on the space coordinates is
         remarkable.  It corresponds to an expansion or stretch
         for $j>0$ and a contraction for $j<0.$  The origin or
         space singularity is unchanged and is fixed.  This is
         shown in Figure \ref{3} for one space dimension for
         positive valued locations for $j=+1$ and $j=-1$ with $k=2n=2.$
         The expansion and contraction for negative valued locations
         is obtained by reflecting the figure through the origin.

           The fixed location of the origin which is a singularity
         in space follows because multiplying it by any scale
         transformation still gives $0$. Also what is remarkable
         is that the immediate neighborhood around the singularity
         serves as an infinite source of points for expansions and
         an infinite sink for points for contractions. This is
         shown schematically in the figure by the appearance of
         unlabelled points in the top figure and the disappearance
         of points in the bottom figure.

         These transformations can easily be extended to $2$ and
         $3$ space dimensions. In this case if the scale
         transformations are the same in each dimension then the
         shape of a figure or object is preserved.  Only its size
         is changed.  However if they are different in different
         dimensions then the shape of a figure is changed.

         The allowed values of time have similar properties to the
         $x$ values shown in Fig. \ref{1}. The values crowd toward
         the accumulation point at $0$. In nonrelativistic quantum
         mechanics where time is simply a parameter for describing
         dynamic evolution, one would expect that all numbers in $R_{n}$,
         including $0$, should be possible values of $t$.  However it will
         be seen later on that  a description of dynamics in terms of
         step operator iteration can never reach $0$. Also time and space are
         treated on an equal footing as coordinates in relativistic
         physics. In this case $t=0$ is
         excluded for the same reasons as were given for
         exclusion of $x=0$ in that the metric
         distance between $0$ and any allowed value of $t$ is not
         a number in $R_{n}$.

         The locations shown in Figures \ref{1}, \ref{2}, and
         \ref{3} also illustrate
         the natural ordering of the points in in each space
         dimension. The ordering is related to arithmetic ordering
         based on iteration of the ordinary $+ 1$ operation
         in arithmetic punctuated by
         exponential jumps.  The relation between the two is given by a
         basic ordering step function $f_{<}$ defined on $R_{n}$  by
        \begin{equation}\label{orderadd}f_{<}(\underline{s}.\underline{t}
        ,2n\underline{e})
        =\left\{\begin{array}{ll}(\underline{s}.
        \underline{t},2n\underline{e})+_{2n}
        (\underline{0}.\underline{0}_{[1,n-1]}1_{n},2n\underline{e}) &
        \mbox{ if }\underline{s}.\underline{t}
        \neq \underline{1}.\underline{1} \\
        (\underline{0}.\underline{0}_{[1,n-1]}1_{n},2n\underline{e}+ 1))
        & \mbox{ if }\underline{s}.\underline{t}
       =\underline{1}.\underline{1}\end{array}\right .\end{equation}
       This definition is for positive $\underline{s}_{[1,2n]}$.
       The subscripts $[1,n]$ have
       been suppressed for brevity. Note  that
       $\underline{e}$ can be positive or negative and is unbounded.
       Also $f_{<}$ is not defined on
       $(\underline{0}.\underline{0},0)$, the $0$ location.

       The inverse operation $f_{>}=f^{-1}_{<}$ can be explicitly
       defined. Or it can be defined
       from $f_{<}$ by
       \begin{equation}\label{invorderadd}f_{>}
       (f_{<}(\underline{s}.\underline{t},2n
       \underline{e})=_{2n} (\underline{s}.\underline{t},2n\underline{e}).
       \end{equation}  Note that
       $(\underline{0}.\underline{0},0)$ is not in the range of $f_{>}$.

       The definition of $f_{<}$ on negative $\underline{s}$ is defined by
       \begin{equation}\label{orderneg}f_{<}(-\underline{s}.
       \underline{t},2n(\underline{e})
       =-(f_{>}(\underline{s}.\underline{t},2n\underline{e})).
       \end{equation} This says that moving
       along with the ordering on negative numbers is equivalent
       to moving opposite to the
       ordering on the positive numbers and changing the sign.

       What is interesting here is that the origin or point
       $(\underline{0}_{[1,n]}.\underline{0}_{[1,n]},0)$
       in any dimension is unreachable by the iterative
       action of  $f_{>}$ on $R_{n}$. Applied to motion
       in a coordinate system, this
       iterative action corresponds to moving backward
       in basic elementary steps
       towards the origin in either the space or time
       directions.  Repeated iterations
       take one arbitrarily close but never reach the origin.

       This property, and the fact that the location is unchanged
       under expansions and contractions reminds one of the big bang
       in cosmology.  The big bang has the property
       that one can go backwards in time and get
       arbitrarily close, but never reach it. Also
       similarities between the expansion
       of the universe and the coordinate system expansion shown
       in Fig. \ref{3} are of interest.\footnote{It is tempting to
       associate galactic black holes with the singularities on the axes
       for some of the degrees of freedom in spaces with two or
       more degrees of freedom. Examples in 2 dimensional space
       are the points with $x=0$ and $y\neq 0$ or conversely.
       These points move under expansions and contractions.  Future
       work will determine if the rigidity of their locations on
       straight line axes can be relaxed.}

      Further development in this direction must await further work.  One should
       note that the proof of the uniqueness of the origin fails
       for $R$ based space and time. It follows that $R$ based
       space and time shows no difference between the origin and
       other points. This is an illustration of the
       possibility that the description of space and time,
       obtained by taking the
       limit $n\rightarrow\infty$, can be different from that
       based directly on $R$ at the outset.

         \section{Quantum Mechanics Based on
         $C_{n}$}\label{QMBCn}

         Perhaps the first point to note about $Qm_{n}$, which is
         quantum mechanics based on $C_{n}$, is that spaces $\mathcal H$ are
         preHilbert spaces\l\cite{Yosida}\r. They are not Hilbert
         spaces as they are not norm complete. This follows from
         the facts that numerical coefficients of states are
         numbers in $C_{n}$ and that the definition of the
         bilinear functional or scalar product $(\psi,\phi)$
         is from pairs of states in $\mathcal H$ to $C_{n}.$

         It is also the case that all operators over $\mathcal H$
         in $Qm_{n}$ are discrete in that their spectra are
         subsets of $C_{n}$.  This follows directly from the
         discreteness of the elements of $C_{n}$.  Thus all
         operators can be expanded as a sum over eigenvalues and
         eigenstates or eigenspaces as $\tilde{A}=\sum_{a}a\tilde{P}_{a}$
         where $a$ is in $C_{n}$ and $\tilde{P}_{a}$ is a projection
         operator over a subspace of $\mathcal H.$

         It is important in discussing states and properties of
         states to keep the various meanings of symbols separate.
         Thus in writing $\psi = c_{a}|a\rangle + c_{b}|b\rangle$,
         $=$ and $+$ have their usual meaning.  But
         the orthonormality of the basis states uses $=_{2n}$ as in
         $\langle a|b\rangle =_{2n}\delta_{a,b}^{n}.$  Here
         $\delta_{a,b}^{2n}=_{2n}1[0]$ if $a=_{2n}[\neq_{2n}]b$ where
         $1$ and $0$ are their representations,
         $(\underline{0}_{[1,n-1]}1_{n}.\underline{0}_{[1,n]},0)$ and
         $(\underline{0}_{[1,n]}.\underline{0}_{[1,n]},0),$
         in $C_{n}.$ Note that for digit strings in $C_{n}$ the
         least and most significant
         subscript labels are $n$ and $1$ respectively. Also as is
         usual the more significant digits stand to the left of
         less significant ones.

         Note that here one takes seriously the use of natural numbers,
         integers and even rational numbers as stand ins or short names for the
         corresponding  numbers in either $C$ or $C_{n}$. Thus $1$
         which appears in the normalization condition is a name in
         $C$ for $1.\underline{0}_{[2,\infty]}=\lim_{\ell\rightarrow\infty}
         1.\underline{0}_{[2,\ell]}.$ Since real numbers can be considered as
         equivalence classes of convergent Cauchy sequences of rational string
         numbers, either $1.\underline{0}_{[2,\infty]}$ or
         $\lim_{\ell\rightarrow\infty}
         1.\underline{0}_{[2,\ell]}$ are names for
         two specific equivalence classes.

         The normalization condition for states such as $\psi =
         c_{a}|a\rangle + c_{b}|b\rangle$ is expressed as
         $$|c_{a}|^{2}_{n}+_{2n}|c_{a}|^{2}_{n}=_{2n}1$$ where
         $|c_{a}|^{2}_{n}\equiv c_{a}\times_{2n}c^{*}_{a}.$ It
         should be noted that in certain cases because of roundoff
         this relation may not be exact.  A good example is the
         state $a|\alpha\rangle +b|\beta\rangle +c|\gamma\rangle$ with
         $a=_{2n}b=_{2n}c$ and $|a|^{2}_{n}+_{2n}|b|^{2}_{n}
         +_{2n}|c|^{2}_{n}=_{2n}(\underline{0}_{[1,n-1]}1_{n}.
         \underline{0}_{[1,n]},0).$
         For $n=2,\; a=_{4} (57.74,-2)=57.74\times10^{-2}$ (decimal) with roundup of the fifth
         figure. With roundup of each square, this gives $1.0002$
         instead of $1.000$.  This type of mismatch is well known
         in computer arithmetic and is why different algorithms
         with different roundoff procedures are used for different
         problems.  It is also why the value of $n$ in
         computations is much larger than the number of
         significant figures in the final outcome.

         Here a specific computer arithmetic is assumed for
         $C_{n}.$ As was the case for the $Th_{n}$, for each arithmetic with its
         specific roundoff prescription and algorithms for
         implementing $+_{2n}$ and $\times_{2n}$ there is a slightly different
         theory $Qm_{n}$. This is not a problem because as $n$
         increases, the number of significant figures needed to
         see the effects of different roundoffs, etc. becomes inaccessible
         to experimental or computational outcomes. Thus one would expect
         that the differences between these theories vanishes in the
         limit as far as agreement or disagreement between theory
         and experiment is concerned.

       $R_{n}$ based space and time is the space and time arena for
       $Qm_{n}$. So position operators and transformation
       operators need to reflect the properties
       of $R_{n}$ space and time.  In each dimension the position operator,
       $\tilde{x}$ has eigenstates $|x_{\pm\underline{s},\underline{e}}
       \rangle$ corresponding to the eigenvalue $x_{\pm\underline{s},
       \underline{e}}$ which is a shorthand notation in $Qm_{n}$ for
       $(\pm \underline{s}_{[1,2n]}.,2n(\underline{e}-1/2))$ as a position value.
       Here $\underline{e}$ includes the sign but not the fixed parameter $n$.

       Expansion of a wave packet state in terms of position states in one
       dimension is given  by \begin{equation}\label{position}\psi =
       \sum_{e=-\infty}^{\infty}\sum_{\underline{s}}=[|x_{+\underline{s},
       \underline{e}}\rangle \langle x_{+\underline{s},
       \underline{e}}|\psi\rangle
       + |x_{-\underline{s}, \underline{e}}\rangle \langle x_{-\underline{s},
       \underline{e}}|\psi\rangle]. \end{equation} The $\underline{s}$ sum is
       over all $\underline{s}$ such that $\underline{s}\neq
       \underline{0}_{[1,2n]}.$ Note that the expansion coefficients are all numbers in
       $C_{n}.$

       It should be noted that the choice made here,
       to exclude singular points from the sum, is arbitrary.
       The reason is that it is unclear
       whether the singular points should be included or excluded
       in such sums over space locations. It is hoped that future work
       will clarify which choice is right.

       This expansion also holds for more than one space
       dimension. In this case one replaces $|x_{\pm
       \underline{s},\underline{e}}\rangle$ by $|x_{\pm
       \underline{s}_{i},\underline{e}_{i}}\rangle$ and sums over
       the dimension label $i$.

       It is tempting to develop $QM_{n}$ further at this point
       and describe motion of systems
       and transformations based on $R_{n}$ space. However,
       at this point in the development, it is
       prudent to defer this to future work.

       \section{Toward The limit $n=\infty$}\label{TLINF}
       It is useful to reemphasize at the outset the importance of
       taking the limit and how it is taken.  At present many
       details of the process are not known.  However the
       requirement that continuum space and time be obtained in the
       limit does guide the process even at this stage of the
       investigation.

       To this end it is useful to look at Eqs. \ref{orderadd},
       \ref{invorderadd}, and \ref{orderneg}. These equations
       and Figs. \ref{1} and \ref{2} show that there are
       regions of $2^{2n}-1$ steps
       of constant spacing punctuated by exponential jumps.
       In one direction the
       spacing intervals are larger by a factor of
       $2^{2n}.$ In the opposite direction they are smaller
       by a factor of $2^{-2n}.$
       As $n$ increases the number of steps and
       size of the regions of constant
       spacing increases exponentially with $n$.
       The exponential jumps become scarcer
       but their size increases exponentially with $n$.

       It is clear from this that if $R$ based space and time and $C$ based
       physics is to be recovered  in the limit of $n=\infty$,
       one must have things arranged so that as $n$ increases the
       positive numbers available to the theory extend from lower and
       upper limits that approach $0$ and $\infty$ respectively as $n$
       increases.  Also there must be no exponential jumps between these limits, and
       the spacing between successive numbers must approach $0$ as $n$
       increases. The same must hold for the negative numbers
       except that the upper limit of $\infty$ becomes a lower
       limit of $-\infty$.

       These conditions are satisfied by the form, $(\pm\underline{s}_{[1,n]}.
       \underline{t}_{[1,n]},\pm 2n(\underline{e}),$ of numbers
       that are used here with $\underline{e}$ set equal to $0$.
       These numbers range from $2^{-n}$ to $2^{n}$ in steps of
       size $2^{-n}$. Jumps described by
       $\underline{e}=0\rightarrow 1$ and
       $\underline{e}\rightarrow -1$ occur at the upper and lower
       limits of the range. For negative numbers the ranges are
       replaced by $-2^{n}$ and $2^{-n}$.

       One sees immediately that these numbers satisfy the desired
       conditions.  As $n$ increases the ranges approach $0$
       and $\pm\infty$ with the spacing going to $0$.  Jumps
       are pushed to $0$ from both sides and out to $\pm \infty$
       in the limit.

       It is easy to see from this discussion why the simpler
       asymmetric form of the numbers, $(\underline{s}_{[1,n]}.,
       n\underline{e})$ are not satisfactory.  The problem is that
       for $\underline{e}=0$ there is an exponential jump right at
       the number $1=(\underline{0}_{[1,n-1]}1_{n}.,0)$.  Also the
       location of this jump  at $1$ is independent of $n$.

       \section{Summary and Outlook}\label{SO}
       Dissatisfaction with the present state of affairs regarding the
       foundational relationship between physics and mathematics forms
       part of the background for this paper. In spite of their obvious
       close relationship they appear to completely unrelated.

       One way to remedy this is to work towards a coherent theory of physics and
       mathematics. Here work in this direction is based on the physical nature of
       language and on the connection between theory
       and experiment.  The physical
       nature of language and numbers leads to the representation of
       language expressions  and numbers as symbol strings that
       are represented by one dimensional physical
       systems in different states.  For numbers this leads to
       their representation as strings of digits in some basis (a
       $k-ary$ representation).

       Regarding the theory-experiment connection it was noted
       that the reality status for very small and very large far
       away objects is more indirect than for nearby laboratory
       sized objects.  This is based on a hierarchal aspect in
       that an experiment test of a theory already implies the
       validity of theories needed to ensure that equipment
       functions properly. The disconnect between theoretical
       predictions as real number solutions to equations and
       measurement outcomes as finite number strings of $n$
       significant figures is noted. The use of computers to
       interpolate between the two different types of numbers is
       noted as is the use of coarse graining to connect
       experimental numbers to real numbers.

       Here a method that ties experiment more closely to theory
       by reducing the "distance" embodied in coarse graining
       (connecting numbers with a few significant figures to numbers
       with an infinite number of figures) is considered.  The
       method consists of replacing the complex numbers on which
       theories are based by complex numbers as pairs of digit
       strings of length $n$.

       The type of number chosen is based on the outputs of
       measurement equipment for measuring continuous quantities
       where the equipment has a threshold sensitivity and
       a finite range of possible outcomes.  These numbers, in the
       form $(\pm \underline{s}_{[1,2n]},2n(\underline{e}-1/2))$ where
       $\underline{s}_{[1,2n]}$ is a $k-ary$ digit string of length
       $2n$, $\underline{e}$ is a $k-ary$ representation of
       integers and $2n(\underline{e}-1/2)$  is an exponent of $k$, correspond
       to $2n$ figure numbers of all possible magnitudes and signs.

       Some properties of space and time and quantum mechanics
       based on these numbers are described. It is seen that
       space and time points are characterized by regions of $k^{2n}-1$
       points with spacing $k^{2n(e-1/2)}$ interspersed with exponential
       jumps to regions of $k^{2n}-1$ points with spacing $k^{n(e(\pm
       1)-1/2)}$.

       An interesting aspect of $R_{n}$ space is the classification
       of points into singular and nonsingular points.
       In $3$ dimensional space the singular points were
       classified into points that had no nearest neighbors in 1
       dimension (coordinate planes), 2 dimensions (coordinate
       axes), and 3 dimensions, (the origin).

       Scale invariant transformations were noted to correspond to
       expansions and contractions of $R_{n}$ based space by
       scale factors that are powers of $k^{2n}$. It was remarkable
       that all points moved under these transformations except
       the origin. Also the origin acts like an infinite source
       and sink for space points for expansions and contractions in all directions.

       A few aspects of $Qm_{n}$, quantum mechanics based on $C_{n}$,
       were described.  The fact that all amplitudes and scalar
       products are required to be numbers in $C_{n}$ was noted.
       Also the need to use computer type arithmetic with roundoff
       to describe normalization of states was outlined.

       The approach to the limit was discussed briefly. It was
       seen that in order to approach the continuum limit of
       $R_{n}\rightarrow R$, the symmetric form $(\underline{s}_{[1,n]}.
       \underline{t}_{[1,n]},2n\underline{e})$of the numbers was
       satisfactory in that as $n\rightarrow\infty$ the region  of
       jump free numbers approached $0$ and $\pm\infty$ for
       $\underline{e}=0.$

       There is much more work needed to expand on the general
       proposal outlined here of describing  theories such as
       $Qm_{n}$ and going to
       the limit $n\rightarrow\infty$.  Energy and momentum and
       many other aspects of
       physics need to be described.  Also more work
       is needed to support the idea that use of the method
       outlined here enables space and time to have different
       properties such as singularities that are invisible to continuum based
       physics, at least as ab initio properties of space time.
       Also number representations
       different from the symmetric one and the one with the $"k-al"$
       point at the end need to be investigated.  Finally the important
       limiting process where one requires that the theory maximally agree
       with experiment for each $n$ must be investigated.

       In conclusion one cannot avoid the speculation that the
       scale transformations of space corresponding to expansions
       and contractions are dynamically driven by forces such as
       gravity and dark energy. This and the possibility of making
       the parameters $n$ and $\underline{e}$ time dependent will
       be investigated in future work.

       \section*{Acknowledgements}
       The author thanks Steve Peiper for discussions on Computer
       Arithmetic. This work was supported by the U.S. Department of Energy,
       Office of Nuclear Physics, under Contract No. W-31-109-ENG-38.


\begin{thebibliography}{99}
\bibitem{Comparith}
             The extent of study can be judged by looking under the subject of
             computer arithmetic on the Amazon web site.
             Over 15,000 book titles are listed.
\end{thebibliography}
            \end{document}